# Protective Molecular Passivation of Black Phosphorous

Vlada Artel[1*], Qiushi Guo[2*], Hagai Cohen[3], Raymond Gasper[4], Ashwin Ramasubramaniam[5], Fengnian Xia[2], and Doron Naveh[1†]

[1] Faculty of Engineering and Bar-Ilan Institute for Nanotechnology and Advanced Materials, Bar-Ilan University, Ramat-Gan, Israel 52900

[2] Department of Electrical Engineering, Yale University, New Haven CT, USA 06511

[3] Department of Chemical Research Support, Weizmann Institute of Science, Rehovot Israel 76100

[4] Department of Chemical Engineering, University of Massachusetts Amherst, Amherst MA USA 01003

[5] Department of Mechanical and Industrial Engineering, University of Massachusetts Amherst, Amherst MA USA 01003

[*] These authors contributed equally to this work

[†] Corresponding author:

DN: doron.naveh@biu.ac.il

**Black phosphorous (BP) is one of the most interesting layered materials, bearing promising potential for emerging electronic and optoelectronic device technologies[1-4]. The crystalline structure of BP displays in-plane anisotropy in addition to the out-of-plane anisotropy characteristic to layered materials[2,5,6]. Therefore, BP supports anisotropic optical and transport responses that can enable unique device architectures[7]. Its thickness-dependent direct bandgap varies in the range of around 0.3-2.0 eV (from single-layer to bulk, respectively), making BP suitable to optoelectronics in a broad spectral range.[8-10] With high room-temperature mobility, exceeding 1,000 $cm^2V^{-1}s^{-1}$ in thin films, BP is also a very promising material for electronics. However, BP is sensitive to oxygen and humidity[11] due to its three-fold coordinated atoms. The surface electron lone pairs are reactive and can lead to structural degradation upon exposure to air, leading to significant device performance degradation in ambient condition.[12,13,14] Here, we report a viable solution to overcome degradation in few-layer BP by passivating the surface with self-assembled monolayers of octadecyltrichlorosilane (OTS) that provide long-term stability in ambient conditions. Importantly, we show that this treatment does not cause any undesired carrier doping of the bulk channel material, thanks to the**

**emergent hierarchical interface structure. Our approach is compatible with conventional electronic materials processing technologies thus providing an immediate route toward practical applications in BP devices.**

The unique and desirable optoelectronic properties of BP[1,3] have motivated much recent work to ameliorate its problematic air and moisture instability. Promising strategies to address this problem include the formation of a protective capping layer or a controlled stable native oxide[15] at the surface. Capping of BP with 2D layered materials such as graphene and hBN[16] have provided stability for a period of 18 days. Alternatively, a 25 nm thick layer of alumina[10] formed by atomic layer deposition was also found to be efficient, especially with the addition of a hydrophobic fluoropolymer layer that improved the stability over several weeks[17]. More recently, organic functionalization of BP with layers of aryl diazonium has been shown to provide effective chemical passivation although this is accompanied by p-doping of the channel material. [18]

Here, we passivate BP with self-assembled monolayers (SAMs) of OTS, thus gaining important advantages in terms of stability, oxidation resistance, and elimination of electronic devices degradation. SAMs are known for their effective surface passivation capabilities, particularly in semiconductor nanostructures.[19,20] OTS molecules, comprised of an eighteen carbon chain backbone, a trichlorosilane head group and a methyl ($CH_3$) functional group, can form smooth and uniform SAMs on oxidized substrates; these crystalline-like, close-packed SAMS [21] can substantially reduce oxygen penetration toward the underlying reactive substrate. A scheme of our device structure is shown in Fig. 1a, where a thin native phosphorus oxide (BPO) layer bridges between OTS molecules and a BP crystal that forms the channel of a field-effect transistor (FET). The source and drain electrodes were defined by electron beam lithography on exfoliated BP flakes followed by metallization (Ti/Au). After liftoff, devices are cleaned with solvents and then immersed in hexane solution of OTS (see methods for details) for one hour to obtain OTS coating on BP.

X-ray photoelectron spectroscopy (XPS) characterization, applied on bulk BP samples, clearly differentiates between the OTS-coated and uncoated BP, both exposed to air for the same period of time. Normalized P 2p core level spectra of coated (blue) and uncoated (red) BP are presented in Fig. 1b from which two pronounced differences are observed. First, in coated samples, the relative intensity of the broad peak around 134 eV, associated with BP oxide (BPO), is significantly smaller than in uncoated samples, indicating a thinner P-oxide layer as compared to bare BP. Because OTS coating of BP was performed in ambient conditions, a very thin layer of BPO still exists at the surface. Second, a pronounced spectral broadening of the low binding-energy line (at ~130 eV) towards higher binding energies is imposed upon the OTS coating. A secondary, smaller broadening of the highly oxidized regime (at ~134 eV) towards lower binding energies is also seen. These spectral signatures of OTS-coated BP are further analyzed in detail and provide a plausible mechanism for the surface reaction, as elaborated upon below.

The efficiency of OTS passivation in preventing the typically rapid degradation of uncoated BP, is further confirmed by Raman spectroscopy, where the intensity of the $A_g^1$ peak of BP (normalized to the intensity of Si substrate peak at 520 cm$^{-1}$) provides a measure of sample's structural stability. In uncoated BP (Fig. 1c) a rapid decrease of the $A_g^1$ signal is observed, associated with the loss of long-range order,[12] due to oxidation and subsequent amorphization of the BP. In contrast, the spectra of OTS-coated samples (Fig. 1d) are stable to the level of fluctuations arising from variations between measurements. Complementary to the Raman analysis, optical microscopy of OTS-coated devices (Fig. 2e-2f) also shows no visible signs of degradation that are typical to BP[11].

The electronic properties probed by transport measurements of FET devices also show clear degradation of uncoated BP devices, with the current response decaying to negligible levels after 14 days (Fig. 2a; also see SI for additional data). Conversely, coated devices maintain their field-effect behavior over a period of 28 days, as shown in Fig. 2b. To ascertain whether the molecular layer dopes the BP, we measured the transport properties of the devices right before and immediately after coating, and found negligible variations in their transport characteristics. In particular, the negligible effect of OTS on threshold voltage (Fig. 2c and Fig. S4) precludes Fermi level shifts or charge transfer between BP and the molecular layer. Furthermore, the on/off current ratios (Fig. 2b and Fig. S3) prove that BP maintains its original electronic properties while stabilized with OTS. These results are significant and, notably, in contrast with other molecular layers such as aryl diazonium on BP[18], they consistently show that BP preserves its original properties under the OTS coating. Fig. 2d further demonstrates that the transconductance (and hence, the hole mobility) is retained during the process of OTS self-assembly. Contrary to the decay of transport with time in bare BP (Fig. 2a) the stability of OTS-coated BP is remarkably persistent, suggesting that a firm corrosion-resistant molecular film is formed on the BP surface.

The success of OTS coatings in stabilizing BP and preserving its original electronic properties was studied in further detail by inspecting the SAM quality and its interaction with BP. Based on XPS (see SI for additional details) we conclude that a nearly vertically-aligned, ordered SAM is formed on the BPO (see Fig. 1a). We find a C:Si atomic ratio (extracted after elimination of background signals) of ~24, which is in excellent quantitative agreement with the value expected after correcting for the standard signal attenuation across the $(CH_2)_{17}CH_3$ backbone of the molecules. More precisely, the theoretical C:Si ratio for a perfect vertical orientation of the molecules is ≈26.4; hence, on average, slightly tilted molecules (~25° to the surface normal) better fit the experimentally-extracted value.

Beneath the OTS layer, a thin (~2nm) BPO layer is detected, consisting of low O-P stoichiometry, with a significant amount of partially-oxidized P components, as shown in Fig. 3c. Importantly, the total signal of oxidized P-2p components (including those components of intermediate oxidation that appear in the coated sample only, as in Fig. 3c) normalized to the bulk BP signal, is roughly conserved in the coated samples with respect to uncoated ones. This fact suggests that limited reduction of pre-formed BPO takes place; a process involving removal of oxygen atoms, while the number of corresponding phosphorus atoms remains unaffected. In fact, the chemical reaction with OTS involves the release of Cl atoms that are expected to cause mild reduction of the top BPO layer[21] analogous to surface oxide reduction and corrosion of Al contacts in electronic devices.[22] Remarkably, the amount of oxygen depleted from the BPO layer is comparable to the amount of oxygen required for the Si-Ox interface between BPO and OTS (see the SI for details). We therefore infer that the binding process of OTS is self-supported by the pre-existing BPO layer, which enables the success and robustness of our treatment.

Our proposed mechanistic interpretation is supported by multiple independent results. Firstly, based on our electrical data we infer that no doping of the BP bulk takes place, which suggests that the bulk BP is indeed unaffected directly by the OTS. In other words, the thin oxide layer that is formed spontaneously, prior to OTS binding, seems to provide efficient protection against BP doping by any of the applied chemical agents. Also, no apparent traces of Cl atoms were observed on the BP surface, indicating that the hydrolysis reaction of OTS was complete. Secondly, controlled surface charging (CSC) data (Fig. S1) shows that BP remains highly conductive. The recorded line-shifts as the electron flood gun (eFG) was switched on and off were about 250 meV for the oxidized P, whereas the BP shifts were as small as 70 meV (see SI for details). As a reference, surrounding regimes of the adhesive tape shifted by ~600 meV under the very same conditions. This analysis suggests that the top BPO layer is slightly reduced during the exposure to OTS, such that the local stoichiometry changes from about $PO_{2.9}$ to about $PO_{1.5}$ on the average. Finally, computed binding energy shifts indicate (as discussed hereafter) a range of intermediate-oxidation states at the BPO/OTS interface that fit well our measured intermediate P-signals in Fig. 3d; signals that are missing from the corresponding P 2p spectrum of uncoated BP.

*ab initio* density functional theory (DFT) calculations were applied as an independent probe of the OTS binding to BP. The structural models considered consist of a unit cell of BPO in a low oxidation state,[15] with a SAM of OTS in either a polymerized or a non-polymerized configuration (Figs. 3a, 3b). From DFT calculations, medium binding-energy states (between 130-133 eV) were obtained, which is in line with the measured spectrum of the OTS-coated BP (Fig. 3c), where signals from partially oxidized states were resolved, as shown in Fig. 3d and further elaborated in the SI tables. The calculated chemical shifts (Tables S3 and S4) roughly fall into two categories: one set of states that are ~0.3-0.6 eV above the bulk P 2p level and a second set at ~1.8-1.9 eV above the lowest P 2p levels. While the actual structure of the surface oxide is most likely amorphous and hence, significantly more complex, the general agreement between core level energies for the surface-functionalized theoretical model and the XPS data supports the picture of chemical bonding, rather than mere physisorption, between the OTS molecules and the BPO.

To conclude, we have demonstrated a simple and effective strategy for efficient, long-term stabilization of BP surfaces against humidity and corrosion using OTS SAMs. This method can be applied to stabilize wafer-scale black phosphorus thin films in the future. In the process of attachment to BP, OTS partially reduces the pre-existing surface oxide, a process quantitatively evaluated by XPS. The stabilization of BP is independently confirmed by transport measurements, Raman spectroscopy and DFT calculations. We have shown that the native oxide layer of BP plays multiple critical roles in the surface functionalization process. Firstly, the BPO layer enables the binding of ordered, close-packed OTS layers by providing the oxygen for the hydrolysis process and presenting a flexible template for assembly. Secondly, the remaining thin oxide layer provides the necessary screening against undesired doping effects associated with inter-diffusion and charge transfer between BP and OTS. Finally, the oxide layer itself is stabilized by OTS and is part of the overall protective capping over the BP channel (see SI for more detail). Overall, this study provides an inexpensive, reliable, and scalable solution to the vexing stability problems of BP thus paving the way for future technological applications.

**Methods**

OTS coating

OTS encapsulation was achieved by immersion of devices in 3ml dry hexane solution containing 30μl of OTS for 1h in a sealed tube to minimize air and moisture exposure. The samples were then washed in cold hexane and soaked for 10 minutes in hot hexane to remove OTS residues. Finally, samples were blow dried with nitrogen.

For coating bulk BP samples (HQ Graphene, 7803-51-2,), the same anchoring procedure was performed, following immersion of BP for one minute in 5% acetic acid solution (in acetone) and blow dried with nitrogen prior to soaking in OTS solution.

Raman spectroscopy

Raman spectra were collected on a Horiba Labram HR spectrometer with exciting laser line at 532 nm.

Device fabrication

BP flakes were mechanically exfoliated on a clean surface of 90nm $SiO_2$/Si wafer. Electrodes were patterned by electron beam lithography and metallized with 3/50nm Ti/Au.

XPS measurements

XPS measurements were carried out on a Kratos Axis Ultra DLD spectrometer, using monochromatic Al kα source at a relatively low power, in the range of 15-75 W. Samples were kept under inert atmosphere prior to their insertion into the vacuum, such that their exposure to air was limited to less than 1 min. The base pressure in the analysis chamber was kept below $10^{-9}$ Torr.

Controlled surface charging (CSC)[23,24] was used in order to differentiate between sample domains, as well as to eliminate signals originated from the underlying adhesive tape (see SI). The CSC data further provides information on the electrical properties of the resolved domains. Complementary measurements were performed on BP-flakes deposited on Ti and exposed to OTS at various conditions (not shown). Curve fitting was done using the Vision software, referring to control measurements on the bare adhesive tape and on samples dominated by polymerized OTS.

DFT calculations

Spin-polarized DFT calculations were performed using the Vienna *Ab Initio* Simulation Package (VASP).[25,26] The core and valence electrons were treated using the projector-augmented wave method.[27,28] The Perdew-Burke-Ernzerhof (PBE) generalized-gradient approximation was used to describe electron exchange and correlation.[29] From convergence studies, a kinetic energy cutoff of 400 eV was chosen. Electronic convergence was accelerated with a Gaussian smearing of 0.05 eV for relaxation calculations; the Blöchl tetrahedron method [30] was used thereafter for single-point electronic structure calculations. Dipole corrections were applied along the vacuum direction of the supercell (normal to BPO layer). Core-level shifts were calculated in the initial-state approximation.

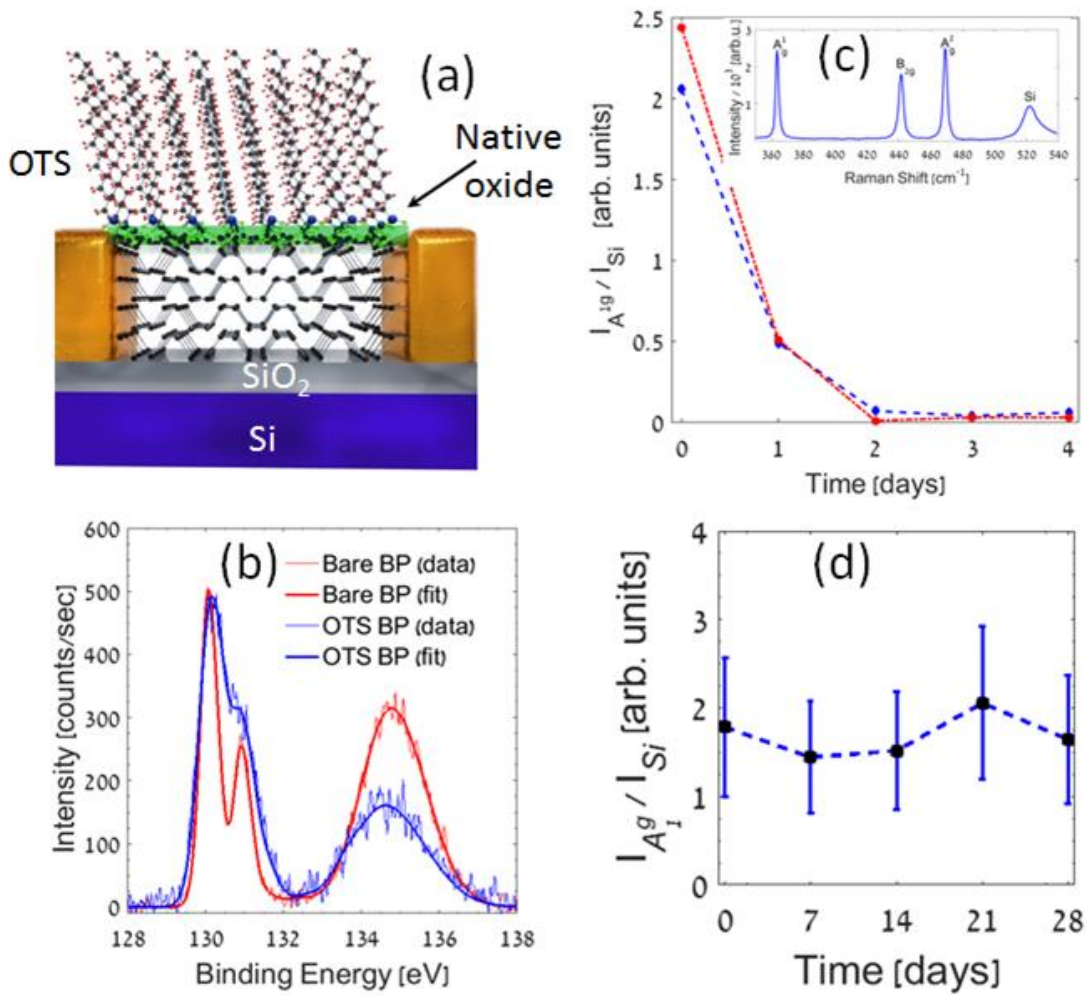

Figure 1. (a) Structure of OTS-passivated BP in a FET device, showing a thin native oxide layer mediating the bonding of BP to OTS; (b) The P 2p XPS spectrum of bare (red) and OTS-coated (blue) BP together with their (red/blue) curve-fitting. Normalized time-resolved Raman spectra showing height ratios between the BP $A_g^1$ peak and the Si substrate peak at 520 cm$^{-1}$ of uncoated (c) and OTS-coated BP (d) samples. A full scanned spectrum is shown in the inset of (c). Colors correspond to different samples in (c).

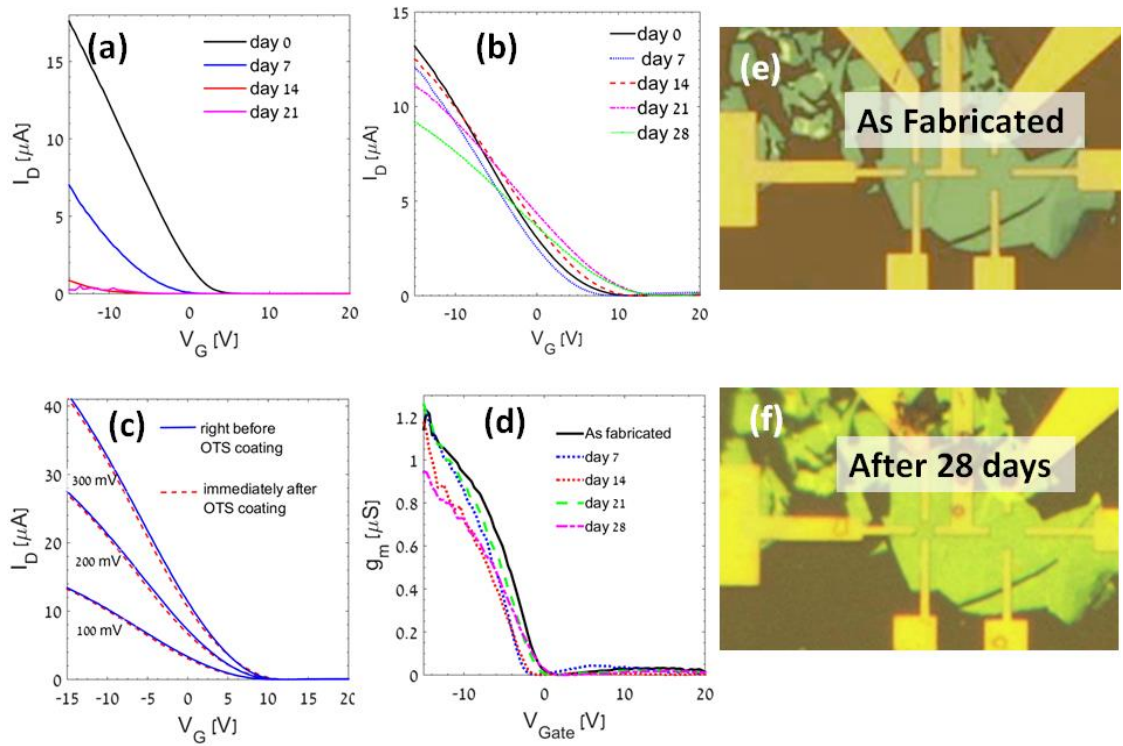

Figure 2. Current transfer curves of uncoated (a) and OTS-coated (b) devices, sampled over a period of 28 days in room atmosphere. I-V curves of as-fabricated (uncoated, solid lines) and right after OTS-coating (dotted curves), showing negligible effect of the passivation treatment on device transfer curves (c), and device transconductrance (d). Optical micrographs of the FET devices before OTS coating (e) and 28 days after OTS-coating (f).

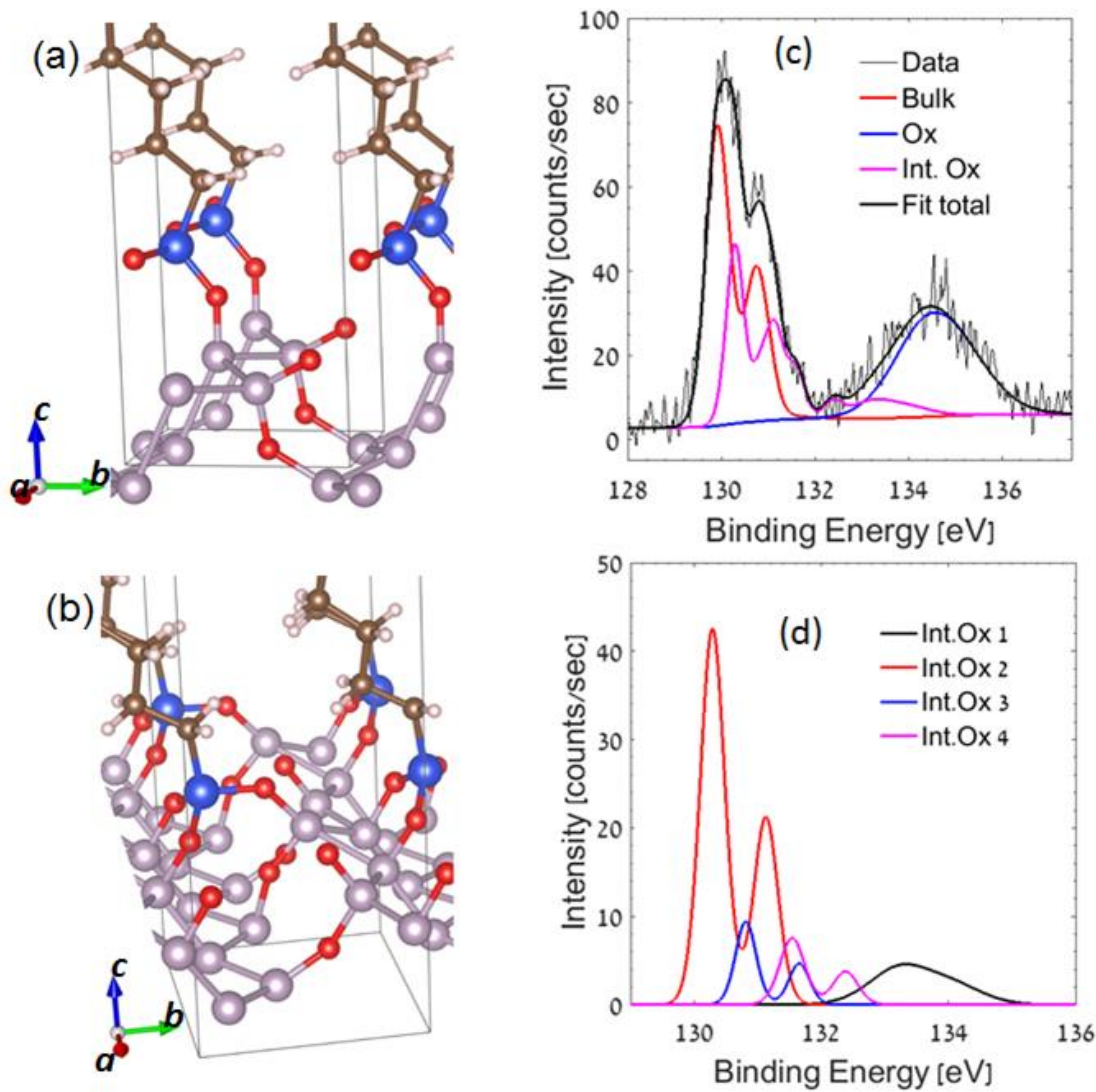

Figure 3. The models used for calculations: A BPO monolayer passivated by OTS where (a) the OTS molecules polymerize (*via* Si-O-Si bonds) along a ridge, and (b) the OTS molecules bind only to surface oxygens without any polymerization. The BPO unit cell in (b) is doubled along the *a*-axis relative to that in (a), such as to accommodate the OTS binding configuration, resulting in ½-monolayer coverage. For clarity, only the part of OTS closest to the BPO layer is shown. (Atom colors: P – grey, O – red, Si – blue, C – brown, H – white). (c) The XPS P 2p line-shape analysis of coated BP, showing the intermediate oxidation states (pink) unique to the coated samples, together with the states common to all samples, bulk BP (red) and native oxide BPO (blue). (d) The four doublet components constructing the curve of intermediate P-states (the pink curve in panel c).

Supplementary Materials for

**Molecular Passivation of Black Phosphorous**

Vlada Artel[1*], Quishi Guo[2*], Hagai Cohen[3], Raymond Gasper[4], Ashwin Ramasubramaniam[5†], Fengnian Xia[2†], and Doron Naveh[1†]

[1] Faculty of Engineering and Bar-Ilan Institute for Nanotechnology and Advanced Materials, Bar-Ilan University, Ramat-Gan, Israel 52900

[2] Department of Electrical Engineering, Yale University, New Haven CT, USA 06511

[3] Department of Chemical Research Support, Weizmann Institute of Science, Rehovot, Israel 76100

[4] Department of Chemical Engineering, University of Massachusetts Amherst, Amherst MA, USA 01003

[5] Department of Mechanical and Industrial Engineering, University of Massachusetts Amherst, Amherst MA, USA 01003

**This PDF file includes:**

Supplementary Text

Figures S1 to S4

Tables S1 to S4

*Details of XPS analysis*

Figure S1 demonstrates the application of CSC (controlled surface charging).[23,24] CSC was used as a means for differentiating between signals of different domains and, to start with, for the elimination of all adhesive tape signals. Second, it was used to exclude differential charging artifacts, thus verifying that the complex P 2p line-shape does indeed reflect various P-oxidation states. The P 2p lines in Fig. S1 were recorded under two markedly different charging conditions and, yet, both samples exhibit small shifts only (much smaller than those encountered at the adhesive tape). Slight differences in shifts characterize the inspected domains: the bulk phosphor, its oxide and the OTS coating. They are summarized in Table S1. It is also noted that the CSC line-shifts provided a useful cross-check of consistency as with the interpretation of atomic concentrations, information that is given in Table S2.

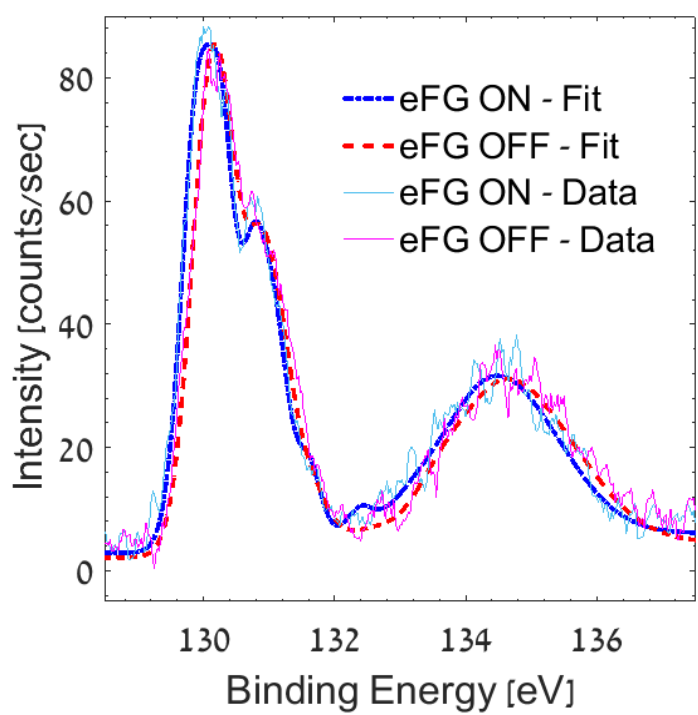

**Figure S1.** XPS P-2p spectra of the OTS-coated BP sample, acquired with (blue dash-dot) and without (red dash) the application of an electron flood gun (eFG) on the sample. Note the limited line shape changes and shifts.

Electrically, the CSC data indicate relatively high conductance of the BP bulk, a feature retained in both samples. Values are given in Table S1. The P-oxide signal manifests slightly larger CSC shifts, which reflects enhanced dielectric character of the very thin overlayer. The shifts of OTS signals in the coated sample, Table S1, are consistent with their nearly vertical organization, as proposed in the paper. Slight deviations can yet be observed, reflecting our limited accuracy in the offline technical work, as well as possible non-uniformities in sample response.

It is noted that straightforward comparisons can be made between elements within each sample. On the other hand, comparing shift values of different samples should be considered more carefully, due to possible variations in the eFG-sample alignment. Specifically, one should be critical as of the measured P-ox shift in the coated sample, found to be larger than in the uncoated one. With this reservation, which in any case deals with minor effects, the observed results are yet in line with the expected behavior, due to additional attenuation and, possibly, extra capturing of eFG electrons at the OTS.

Testing the subsequent exposure of the samples to air (~ 24 hours, not shown) resulted in increased oxidation in the uncoated samples, while the coated ones exhibited a significant, though not perfect, oxide-protective feature. Quantitatively, we find it very likely that the initially formed P-oxide serves as the main source of O-atoms that participate in the bonding of OTS molecules (via Si-O-P bonds), as well as for the evolution of lateral polymerization at the OTS (via Si-O-Si bonds). Due to signal overlapping, the level of our quantitative analysis was limited; hence, further study is needed to improve our understanding of the OTS-BP interface details.

**Table S1**: The CSC-induced line-shifts (in meV), obtained upon on/off switching of the electron flood gun (eFG). Unless otherwise mentioned, the experimental error is below ±20 meV. For those cases encountering non-uniform shifts, a range of values is given.

|  | P-bulk | P-ox |  |  | OTS |  |  |  |
| --- | --- | --- | --- | --- | --- | --- | --- | --- |
|  | P | $P_{int}$ | $P_{ox}$ | $O_{phosphorus}$ | Si | $C_{main}$ | $C_{Si}$ | $O_{OTS}$ |
| **Coated** | 70 | 100-400 | 220 | 260 ±25 | 225 | 275 | 245 | 210 ±100 |
| **Uncoated** | 70 | --- | 145 | 155-180 |  |  |  |  |

**Table S2**: Atomic concentration ratios, normalized to the amount of bulk-P, as evaluated after elimination of the adhesive tape signals. Raw binding energies (in eV) are given as well, extracted under eFG-off conditions. The $C_{Si}$ signal corresponds to the C-atom bonded to Si.

|  | P-bulk | P-ox |  |  | OTS |  |  |  |
| --- | --- | --- | --- | --- | --- | --- | --- | --- |
|  | P | $P_{int}$ | $P_{ox}$ | $O_{phosphorus}$ | Si | $C_{main}$ | $C_{Si}$ | $O_{OTS}$ |
| **Coated** | 1 | 0.73 | 0.74 | 2.60 | 1.09 | 26.3 | 1.06 | 1.10 |
| $E_B$ (off) | 130.02 | 130.38 – 133.20 | 134.57 | 531.96 & 533.48 | 102.37 | 285.70 | 283.98 | 532.24 |
| **Uncoated** | 1 | --- | 1.43 | 4.20 |  |  |  |  |
| $E_B$ (off) | 130.06 |  | 134.57 | 531.94 & 533.51 |  |  |  |  |

*Density Functional Theory Calculations*

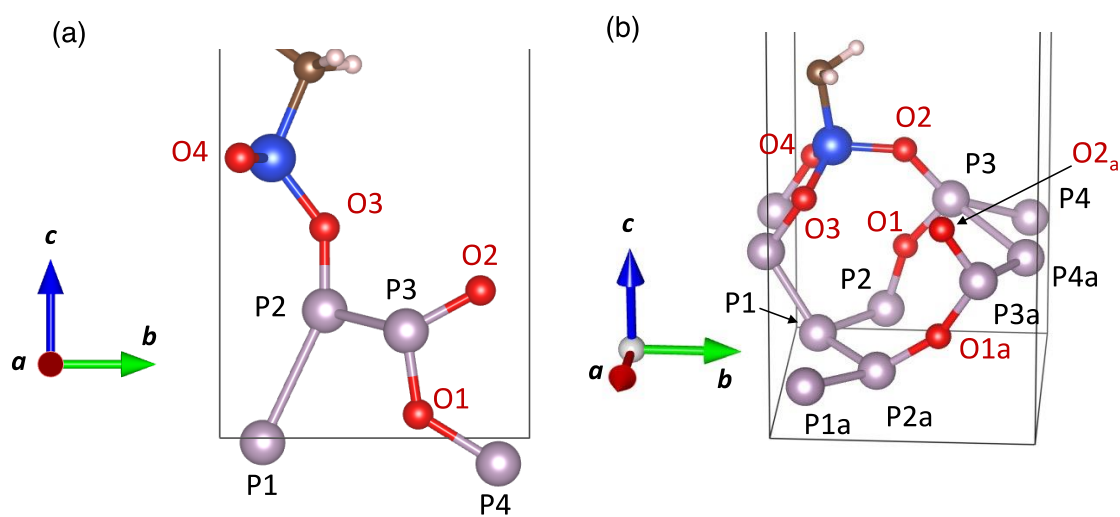

**Figure S2**: Structural models of BPO monolayer passivated by (a) polymerized and (b) non-polymerized OTS molecules. The labels for the P and O atoms correspond to those in the core-level shift data in Tables **S3** and **S4.** Only the OTS head group is retained here for clarity. (Atom colors: P – grey, O – red, Si – blue, C – brown, H – white)

**Table S3**: 2p core level (CL) energies (in eV) and differences relative to lowest core level ($\Delta_{CL}$) of phosphorus atoms in BPO with both polymerized (Poly-) and non-polymerized (Nonpoly-) OTS passivation (see Fig. S2 for indexing of atoms)

| Atom | Poly-OTS | | Nonpoly-OTS | |
|---|---|---|---|---|
| | CL | $\Delta_{CL}$ | CL | $\Delta_{CL}$ |
| **P1** | -124.97 | 0.48 | -124.79 | 0.67 |
| **P1a** | - | - | -124.85 | 0.61 |
| **P2** | -125.45 | 0 | -125.46 | 0 |
| **P2a** | - | - | -125.10 | 0.36 |
| **P3** | -125.19 | 0.26 | -125.00 | 0.46 |
| **P3a** | - | - | -124.97 | 0.49 |
| **P4** | -123.89 | 1.56 | -123.61 | 1.85 |
| **P4a** | - | - | -123.63 | 1.83 |

**Table S4**: 1s core level (CL) energies (eV) and differences relative to lowest core level ($\Delta_{CL}$) of oxygen atoms in BPO with both polymerized (Poly-) and non-polymerized (Nonpoly-) OTS passivation (see Fig. S2 for indexing of atoms)

| Atom | Poly-OTS | | Nonpoly-OTS | |
| --- | --- | --- | --- | --- |
| | **CL** | **$\Delta_{CL}$** | **CL** | **$\Delta_{CL}$** |
| **O1** | -511.92 | 0.12 | -511.89 | 0.06 |
| **O1a** | - | - | -511.95 | 0 |
| **O2** | -510.09 | 1.95 | -511.58 | 0.37 |
| **O2a** | - | - | -510.05 | 1.9 |
| **O3** | -512.04 | 0 | -511.72 | 0.23 |
| **O4** | -509.74 | 2.3 | -511.68 | 0.27 |

**DFT Methodology**

The initial BPO structure was based on the work of Edmonds *et al*.[7] Atomic positions and in-plane cell dimensions were relaxed with a force tolerance of 0.02 eV/ Å. All systems had approximately 20 Å of vacuum normal to the BPO layer to prevent spurious interaction between images. For relaxation calculations, the Brillouin zone was sampled with a 4×3×1 Γ-centered *k*-point mesh for the polymerized OTS case and a 2×3×1 Γ-centered *k*-point mesh for the non-polymerized case; this leads to approximately the same *k*-point density in both cases. The core level shifts of the BPO monolayer upon OTS adsorption were calculated using the initial state approximation, and referenced against the vacuum level. As all cases require an appreciable dipole correction, the average of the two vacuum levels was used as the reference for electronic energies. Core level shift calculations for all systems were performed at higher accuracy using a 600 eV kinetic energy cutoff and double the *k*-point mesh density.

*Electronic stability of OTS-BP devices*

Transfer curves of OTS-coated devices acquired over 28 days from fabrication show relatively small deviations. In a stringent comparison, the passivated BP devices prove to even outperform the stability of bare $MoS_2$ FET devices.[8] Oxidation of MoS2 typically require high temperature and plasma to drive and enhance the reaction.[9,10] The I-V transfer curves of two devices are shown in Fig. S3, establishing the consistency of the results shown in Fig. 2a.

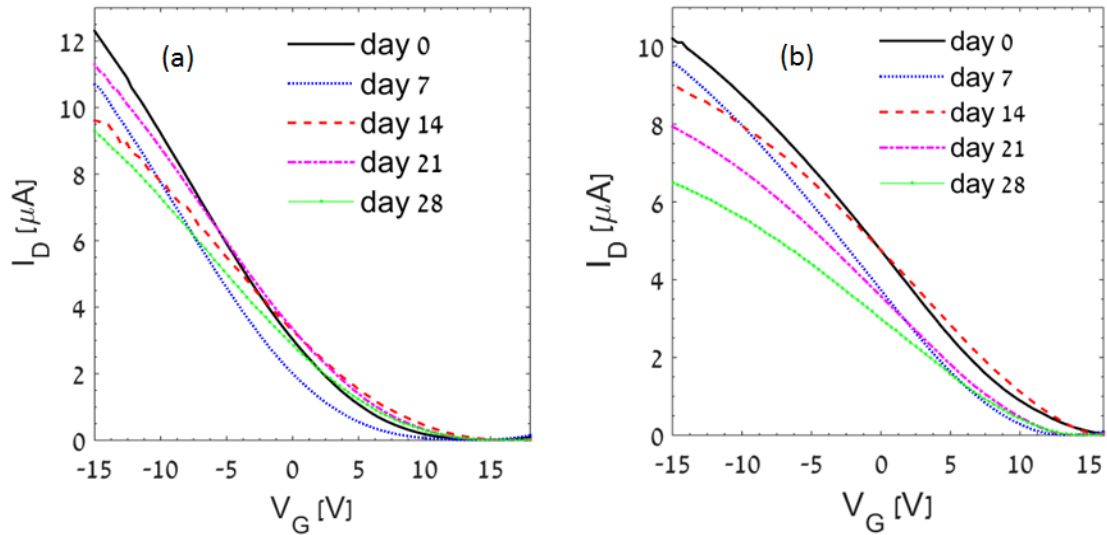

**Figure S3.** I-V transfer curves similar to those of Fig. 2a, acquired from two additional OTS-passivated BP devices.

Additional to the transconductance shown in Fig. 2d, the stability in performance of the passivated devices is summarized in the trends of threshold voltage of the passivated devices, as in Fig. S4.

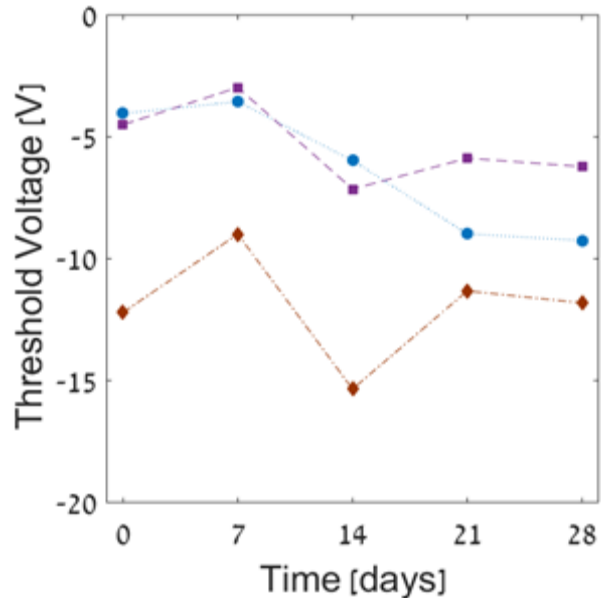

**Figure S4.** A summary of the measured threshold voltage of three BP-OTS devices over 28 days in ambient.